# Information Transfer and Landauer's Principle

**Michael C. Parker**

Fujitsu Laboratories of Europe, 38 Byron Avenue, Colchester, CO3 4HG, UK.
M.Parker@ftel.co.uk

**Stuart D. Walker**

University of Essex, Department of Electronic Systems Engineering, Wivenhoe Park, Colchester, Essex, CO4 3SQ, UK.
stuwal@essex.ac.uk

**Abstract:** In this paper we present an analysis of information transfer based on Landauer's principle (i.e. erasure of information is associated with an increase in entropy), as well as considerations of analyticity and causality. We demonstrate that holomorphic functions allowing complete analytic continuation cannot propagate any information, such that information transfer only occurs with analytic functions having points of non-analyticity (i.e. meromorphic functions). Such points of non-analyticity (or discontinuities) are incompatible with adiabaticity, so that information transfer must always be accompanied by a change in entropy: a dynamic reformulation of Landauer's Principle. In addition, since Brillouin proved that discontinuities cannot travel faster than the speed of light $c$, this also implies that information cannot be transferred at superluminal speeds.

## 1. Introduction

Recent reports of "fast light", e.g. [1], have excited interest due to the implications for superluminal data transmission. However, the theoretical analyses associated with these reports have neglected to include Landauer's principle [2] and have not fully considered the relationship between entropy and information in a complete information-thermodynamic treatment. The close association of information and entropy has already been well described [3,4], often in negative correlation, e.g. Brillouin's concept of negentropy (i.e. negative entropy) for information [5]. In the following analysis, we will demonstrate that proper inclusion of these information-thermodynamic principles yields both a more rigorous definition of what we mean by information, and leads to a full retention of classical causality.

Landauer's principle [2], in essence, says that the erasure of information requires energy, and hence is associated with an increase in entropy; whilst in contrast, information creation does not require energy, and is not associated with a change (e.g. decrease) in entropy. This is a 'static' statement, since it only relates to the erasure/creation of information at a particular spatial position. In this paper, we apply Landauer's principle to the 'dynamic' situation of information transfer, where information is moved or translated from one spatial position to another. Hence, the mathematical arguments and discussions in this paper are directed towards the idea that the transfer of information from one spatial position A to another spatial position B is associated with a change (i.e., in general, an increase) in entropy.

Sommerfeld and Brillouin pioneered the use of a step-discontinuity to analyse group velocity [6], and found that in any causal medium a step-front travels at the speed of light in vacuum $c$, via forerunners, such that causality is not compromised. Propagation of truncated Gaussian or cosine pulses in a dispersive medium has also been extensively analysed, e.g. [7-12]. In contrast to the step-discontinuity, such pulses allow the process of analytic continuation. It has been suggested that a complete temporal/frequency description of the pulse may be found in the leading edge of the pulse. Thus a replica of the complete pulse can be 'carved' out of the energy pedestal of the leading edge (forerunner travelling at $c$) of the pulse and allowed to propagate. Comparing the spacetime positions of the centroids of the initial pulse and the transmitted (*i.e.* replica) pulse appears to indicate a superluminal velocity. This is an inappropriate comparison, however, since the replica pulse was already present in the leading edge of the pulse (which precedes the centroid), and which travels at $c$. Hence any information associated with the pulse was already in the forerunner that travels at a speed $c$ (or less), and physical superluminal velocities do not occur.

**2. Differential Information and Entropy Change**
*2.1 Analytic Functions & Filter Theory*
Complex calculus theory has been extensively applied to the problem of group velocity determination, e.g. [6]. However, the properties of analytic functions have not to our knowledge been explicitly applied to information transfer. These functions have both local and global (i.e. distributed) properties, and obey the Cauchy-Riemann equations. An analytic function that contains isolated poles in the complex plane is known as meromorphic [13]. Analytic continuation can be used to reconstruct a meromorphic function over the whole complex plane, except for the points of non-analyticity (poles), where the Cauchy-Riemann equations no longer apply. Hence, when a signal used to 'transmit' information from A to B allows analytic continuation between those two points, no information transfer takes place between A and B, since such a signal is already completely defined at the destination point B.

In an important paper by Toll [14], it was found that a function, which is causal (i.e. non-anticipative) in time, has an analytic Fourier transform (FT). Its frequency components in the complex frequency plane are analytic and obey the Cauchy-Riemann equations. We note that a sufficient condition for the analyticity of a function is that it obeys the Cauchy-Riemann equations [15]. An analytic function allows analytic continuation at any point in the complex plane where the Cauchy-Riemann conditions hold. Generalising Toll's paper, any function that is zero from a fixed point back to minus infinity on the real axis (*i.e.* a 'spatial' causality or boundedness) will have an analytic FT. This means that any 'windowed' or 'apertured' function has an analytic FT. Hence, a Young's twin-slit interference pattern, for example, is analytic and allows for analytic continuation. Likewise, reflection and transmission coefficients are given by the FT of the finite spatial distribution of scatterers (bounded in spatial extent) [16], such that they too are analytic. Thus both transverse and longitudinal grating structures scatter light analytically. Since the Maxwell equations are also an example of the Cauchy-Riemann equations [17], solutions to equations describing wave propagation through finite structures are therefore analytic.

Filter theory has also been extensively developed via analytic function theory to describe physically realizable network transfer functions. Two important tenets are concerned with causality (non-anticipative functions) and stability. The first of these is satisfied by the Paley-Wiener criterion [18,19], whilst stability is ensured if the filter function's denominator is a Hurwitz polynomial [18]. We now show (we believe for the first time) that these criteria must also define physically realizable information-bearing signals. We point out that the Gaussian function does not satisfy the Paley-Wiener criterion, whilst standard trigonometric functions fail the Hurwitz requirement by having poles in the lower-half of the complex plane.

*2.2 Differential Information and Complex Function Theory*
If we are to have a meaningful concept of information transfer, and therefore associated spatial differences in information, that aspect can only be described by the differential information. The differential information (in units of bits) of a continuously-varying function $p(x)$ in space $x$ is given by [20]:

$$I = \int_{-\infty}^{\infty} p(x)\log_2 p(x)dx \qquad (1a) \qquad \text{where} \qquad \int_{-\infty}^{\infty} p(x)dx = 1. \qquad (1b)$$

In the infinitesimal limit of the well-known discrete summation describing information, there is a diverging part that we ignore in (1a). This is because when considering differences between differential informations the diverging parts cancel; the constant (infinite) divergent part due to the infinitesimal limit being the same everywhere. Since we are comparing the relative information between two spatial positions, A and B, comparison of the differential information at each of these two locations is therefore the meaningful metric. We assume that $p(x)$ is the intensity distribution of a wave amplitude distribution $\psi(x)$ satisfying a wave equation for a physically realistic system, such that $p(x) = |\psi(x)|^2$. Equation (1b) assumes that $p(x)$ is normalized, so that it is also equivalent to a probability distribution. As indicated in section 2.1, if $\psi(x)$ is a solution to a wave equation for a finite scattering medium, it and its complex conjugate allow analytic continuation. We assume that in general $\psi(x)$ can be expressed as a sum of two purely-real functions of $x$, such that $\psi(x) = r(x) + is(x)$. Hence by analytic continuation we can employ the function $A\psi(x) = g(x,y) + ih(x,y)$, where $A$ is the analytic continuation operator on $\psi(x)$, such that $x$ is replaced by $z = x + iy$. The complex plane $z$ is given by $z = x + iy$, $y$ being the imaginary-axis co-ordinate, whilst the purely-real functions $g(x,y)$ and $h(x,y)$ obey the Cauchy-Riemann equations $\partial_x g = \partial_y h$ and $\partial_y g = -\partial_x h$. Together, $g$ and $h$ form an analytic function, and whilst they are different to the functions $r$ and $s$, along the real $x$-axis where $y = 0$ we have that $g(x, y=0) + ih(x, y=0) = r(x) + is(x)$. We note that the complex conjugate of $A\psi$, given by $[A\psi]^* = g(x,y) - ih(x,y)$, is not an analytic function, since it does not obey the Cauchy-Riemann equations. However, the complex conjugate of the wave amplitude $\psi^*(x)$ given by $\psi^*(x) = r(x) - is(x)$ can be analytically continued to yield $A\psi^*(x) = u(x,y) + iv(x,y)$, where again the purely-real functions $u$ and $v$ obey the Cauchy-Reimann equations, and form an analytic function. Along the real $x$-axis (where $y = 0$) we again have that $u(x, y=0) + iv(x, y=0) = r(x) - is(x)$. In general, the functions $u$ and $v$ are different to $g$ and $h$, respectively, since the combined operations of analytic continuation and complex conjugation do not commute (and also differ in whether the resulting function is analytic or not). We rewrite equation (1a) as:

$$I = \int_{-\infty}^{\infty} \psi(x)\psi^*(x)\log_2\left[\psi(x)\psi^*(x)\right]dx. \qquad (2)$$

The products of analytic functions, and power series of analytic functions are also analytic [13], hence the integrand in equation (2) can be analytically continued to yield a closed contour integration in the complex $z$-plane:

$$\oint (g+ih)(u+iv)\log_2\left[(g+ih)(u+iv)\right]dz = 2\pi i \sum R \qquad (3)$$

where $\sum R$ is the sum of the enclosed residues, from the Cauchy residue theorem. In the conventional manner, we can express the closed contour integral as an integral along the real $x$-axis, followed by an anti-clockwise semicircular line integral $\Omega$ in the upper-half complex $z$-plane, and let the range along the $x$-axis and the radius of $\Omega$ tend to infinity, so as to yield:

$$\int_{-\infty}^{\infty} p(x)\log_2 p(x)dx + \int_{\Omega}(g+ih)(u+iv)\log_2\ (g+ih)(u+iv)\ dz = 2\pi i \sum R. \qquad (4)$$

From the Paley-Wiener criterion we must have that $g,h,u,v \to 0$ as $|z| \to \infty$. In which case, the semicircular line integral along $\Omega$ will equal zero. Hence the information associated with a function $p(x)$ is given by the sum of the associated residues:

$$I = \int_{-\infty}^{\infty} p(x)\log_2 p(x)dx = 2\pi i \sum R \ . \tag{5}$$

Residues are due to singularities and poles, where analytic continuation cannot be performed (and the Cauchy-Riemann equations do not locally hold), such that equation (5) indicates that the differential information of a function is therefore associated with these points of discontinuity. A holomorphic function allowing analytic continuation across the whole of the complex plane does not have any points of discontinuity, such that the sum of the residues is zero. Hence, according to equation (5) a holomorphic wave function contains zero differential information, and therefore cannot be used to transmit information anywhere. Only meromorphic functions (which have points of non-analyticity) have differential information associated with them. The information associated with a physically realizable meromorphic (information-bearing) function is thus simply given by the Cauchy residue theorem of equation (5). We believe this to be a novel application of the calculus of residues to information theory. An information-bearing signal must therefore have points of non-analyticity [21,22], so as not to allow complete 'prediction' or analytic continuation in the complex plane. We note that a point of non-analyticity cannot be a zero-order discontinuity, since that has finite power at infinite frequencies and hence does not fulfill the Paley-Wiener criterion [23]. Since information is associated with points of non-analyticity (i.e. discontinuities), Brillouin's work [6] shows that superluminal information propagation is not possible.

*2.3 Information Transfer and Entropy Change*
Landauer's principle requires the erasure of information to be associated with an increase in entropy; whilst creation of information is not associated with a change (e.g. reduction) in entropy. Hence, transfer of information from A to B can be understood to be the erasure of the information at location A (which from Landauer's Principle is accompanied by an increase in entropy), proceeded by the re-creation of that same information at B. Thus transferring information is associated with an increase in entropy. Likewise, the points of non-analyticity and discontinuities associated with information are inimical to assumptions of 'smoothness' and adiabaticity in a dynamic system. Hence, information transfer must be accompanied by changes in entropy: a dynamic reformulation of Landauer's principle. Finally, we note that information transfer is distinct from cloning or copying of information from A to B. This action cannot be perfectly performed due to the no-cloning theorem [24], and the impossibility of noiseless amplification. Both of these processes are related to information loss and hence entropy increase; the main thrust of this paper.

## 3. Conclusions
In this paper we have demonstrated that information can only be carried by non-holomorphic functions. The points of non-analyticity associated with these functions are points of discontinuity, such that adiabatic conditions cannot be maintained for information transfer. Thus information transfer must be accompanied by a change in entropy: a dynamic reformulation of Landauer's principle. Points of non-analyticity and discontinuities, i.e. information, cannot travel faster than *c* in any medium, thus maintaining causality.